\documentclass[instruments,article,submit,pdftex,moreauthors]{mdpi} 
\usepackage{upgreek}
\firstpage{1} 
\pubvolume{1}
\issuenum{1}
\articlenumber{0}
\pubyear{2026}
\copyrightyear{2026}
\datereceived{ } 
\daterevised{ } % Comment out if no revised date
\dateaccepted{ } 
\datepublished{ } 
\Title{Design and operation of a flash lamp for vacuum ultraviolet light production}

\Author{Silas Bosco$^{1}$, Jonas Bürgi$^{1}$\orcidH{}, Livio Calivers$^{1}$\orcidA{}, Richard Diurba$^{1}$\orcidE{}, Johannes Furrer$^{1}$\orcidK{}, Jan Kunzmann$^{1}$\orcidF{}, %Lorenzo Meier$^{1}$, 
Saba Parsa$^{1}$\orcidC{}, Sascha Rivera$^{1}$\orcidL{}, Nicolas Sallin$^{1}$\orcidJ{}, %Alexander Schait$^{1}$, 
Camilla Tognina$^{1}$,
Serhan Tufanli$^{1}*$,
Michele Weber$^{1}$\orcidB{},
Dominik Wermelinger$^{1}$\orcidI{}}

\AuthorNames{Silas Bosco, et. al.}

\address{%
$^{1}$ \quad Laboratory for High Energy Physics, Albert Einstein Center for Fundamental Physics, Universität Bern, 3012 Bern, Switzerland
}

\corres{Correspondence: serhan.tufanli@unibe.ch}

\abstract{Noble liquids, notably argon and xenon, are utilised as both detector media and as the detector target for dark matter and neutrino physics experiments. When the noble liquid is excited by particles, it scintillates vacuum ultraviolet light, which sensors then detect. A major focus of the detector development community is on producing precision light sensors for noble liquid detectors. We introduce a flash lamp to test VUV-sensitive light sensors with light at wavelengths observed at noble liquid detectors. This paper discusses the design and presents results from a flash lamp prototype operated at room temperature.}

\keyword{noble liquid scintillation light, detector development}

\begin{document}

%%%%%%%%%%%%%%%%%%%%%%%%%%%%%%%%%%%%%%%%%%
\section{Introduction}
Argon and xenon-based detector technologies are utilised by many neutrino and dark matter experiments~\cite{DUNE:2020lwj,DUNE:2021tad,DEAP-3600:2017ker,DarkSide-20k:2017zyg,XENON:2024wpa,LZ:2019sgr}. The excitation of the detector medium by ionising particles generates prompt scintillation light via argon and xenon excimer de-excitation, yielding photons in the vacuum ultraviolet (VUV). For argon, this light has a peak wavelength at 128~nm~\cite{Doke_2002}. Experiments surround the detector material with light sensors to maximise signal detection. The light detection signals from the sensors are then used to identify and characterise particles in the medium. 

Direct tests of VUV-sensitive light sensor performance in noble liquids reproduce the desired spectrum but are time- and infrastructure-intensive as they require cryogenic systems. Test setups for VUV-sensitive light sensors rely on light-emitting diodes (LEDs), whose spectra do not extend into the VUV. Improvements have been made to provide VUV light without significant infrastructure at relevant wavelengths with deuterium lamps, such as Nikhef's VULCAN system~\cite{marjolein}. However, these commercial lamps produce spark emission and quasi-constant light, which is not ideal for signal-shaping or single-photoelectron performance studies. These limitations highlight the need for a room-temperature device that promptly produces VUV photons with high spectral purity. An example of such an argon lamp is presented in~\cite{McDonald:2021qec}.

We have designed a new argon flash lamp. It is a compact apparatus that exploits electrical discharges in argon gas to generate VUV light through the same excimer de-excitation processes occurring in the liquid phase. The lamp is intended to be portable and flexible to changes in electrode distance and gas pressure, while enabling precise, repeatable characterisations of photosensors for noble element detectors. The testing setup proposed here also produces light pulses on the microsecond-timescale to allow the data acquisition system to synchronise timing between the lamp and external light sensors. Unlike previous flash lamps, it utilises a filter for VUV light and includes a new capability to adjust the distance between the electrodes. Additionally, it does not require optical instrumentation and only uses a vacuum pump and gas supply, providing a simpler design suitable for multi-institutional testing in the context of high energy physics experiments. This paper describes the design and operation of the flash lamp. 

%%%%%%%%%%%%%%%%%%%%%%%%%%%%%%%%%%%%%%%%%%
\section{Design of the Argon Flash Lamp}
A chamber with sparking copper electrodes is the core of the argon flash lamp. The chamber is a cylindrical vessel 65~mm in diameter and 73~mm tall, machined from poly-ether-ether-ketone (PEEK). This high-performance thermoplastic offers exceptional electrical insulation, chemical inertness, and mechanical strength. Its low outgassing rate and dimensional stability across a broad temperature range make PEEK especially well-suited for vacuum and high-voltage environments, permitting precise electrode alignment and reliable operation through repeated discharge cycles.

\begin{figure}[t!]
    \centering
    \includegraphics[width=0.48\linewidth]{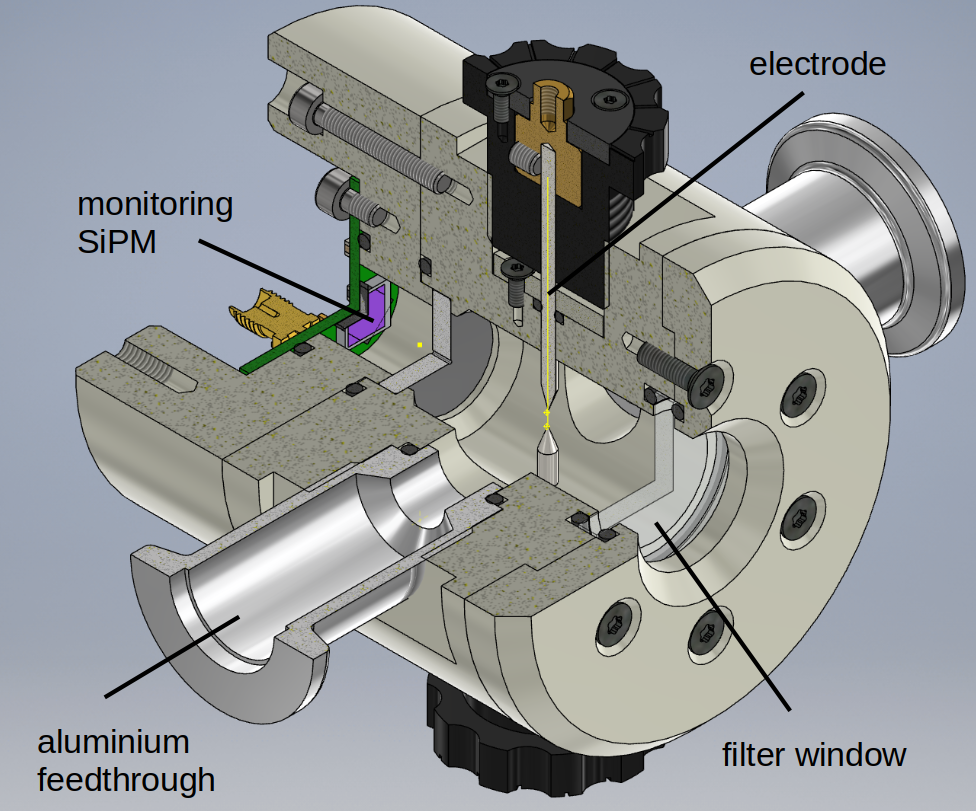}
    \caption{Diagram of the flash lamp chamber with an interior view of the electrodes and filter.}
    \label{fig:chamber}
\end{figure}

\begin{figure}[H]
    \centering
        \includegraphics[width=0.4\linewidth]{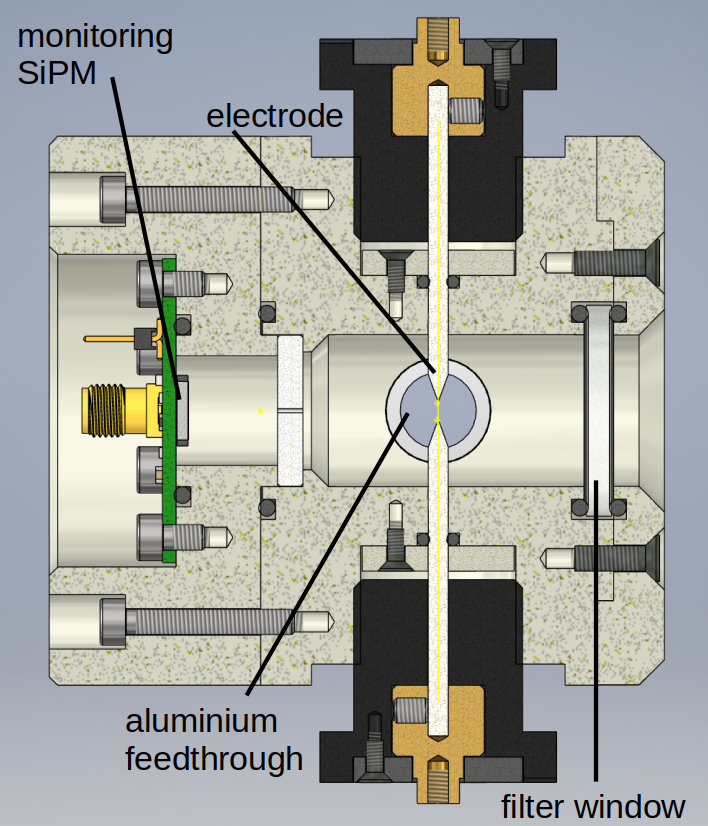}
    \includegraphics[width=0.4\linewidth]{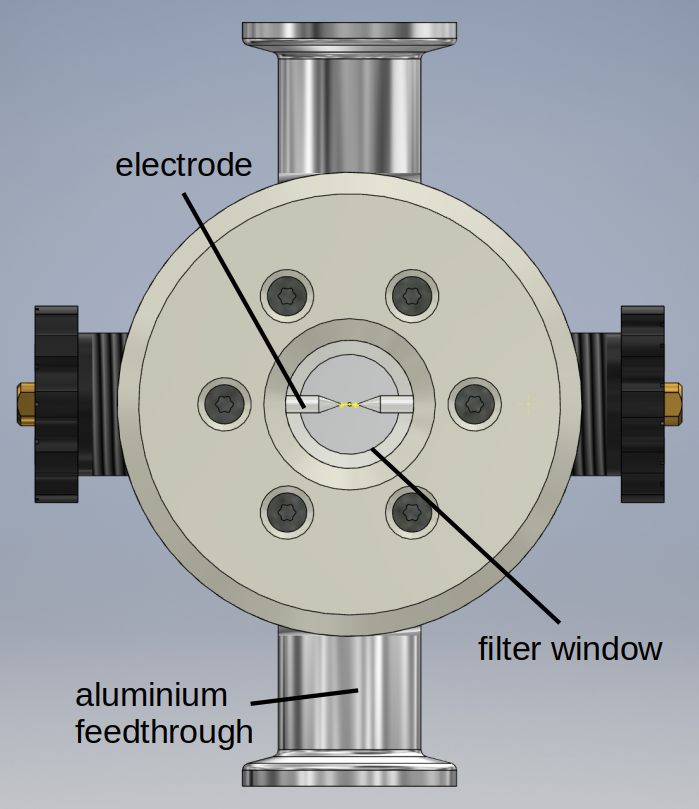}
    \caption{Diagram of the flash lamp chamber, showing a cross-sectional view from the side (left) and the front (right) of the device.}
    \label{fig:chamber_front}
\end{figure}

A diagram of the chamber is shown in Figure~\ref{fig:chamber}, with Figure~\ref{fig:chamber_front} showing diagrams from the side and front views of the chamber. Two aluminium feedthroughs are integrated into the sidewall. High-purity argon from a bottle of 99.998\,\% argon travels through these aluminium feedthroughs. The chamber is evacuated and refilled with argon every few minutes. VUV light flashes are produced between a pair of opposing electrodes positioned at the top and bottom of the chamber and aligned along the central axis, as shown in Figure~\ref{fig:chamber_front}. The copper electrodes can be in contact, and the rotary mechanism allows for their distance to increase in 0.1~mm increments to a maximum separation of 10~mm. The internal parts of the rotary adjustment mechanism are fabricated from copper to ensure excellent conductivity. The assembly is sealed with O-rings to maintain vacuum integrity while providing smooth rotation and electrical isolation from the chamber body.

A monitoring SiPM connected to a printed circuit board (PCB) is installed on one side of the chamber. This SiPM is inside the chamber to monitor flash synchronisation and light production. On the opposite side of the chamber from the monitoring SiPM, a filtering window is covered with a P/N 25125FNB VUV-Optical Filter produced by eSource Optics~\cite{optical_window}. The filter transmits light within a peak of 125~nm $\pm$ 2.5~nm and a full width of 20~nm $\pm$ 5.0~nm, ensuring emission from argon excimer decays is transmitted. The light travels through the filtering window to exit the chamber. Only $10^{-3}$ to $10^{-4}$ of light outside of the filter range is transmitted, significantly reducing non-VUV light emitted from the chamber.

\begin{figure}[H]
    \centering
    \includegraphics[width=0.95\linewidth]{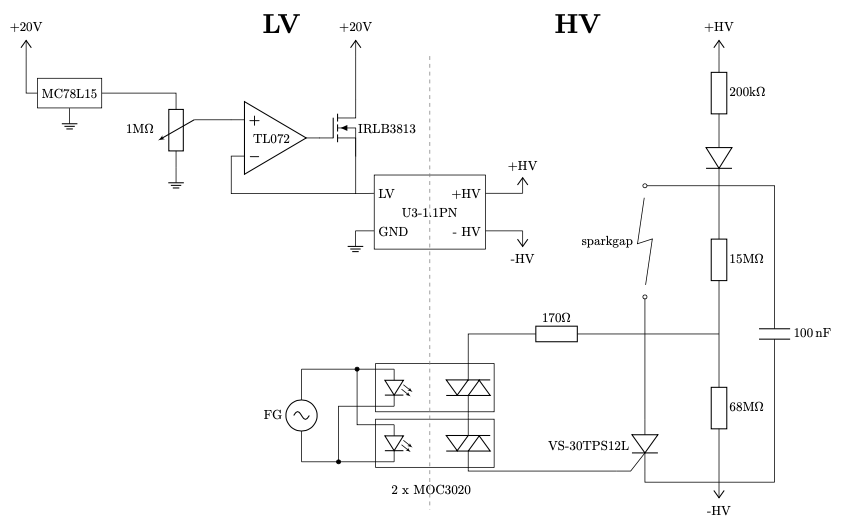}
    \caption{Diagram of the electrical circuit for the spark generation. The circuit consists of two parts: the first part is the low voltage (LV), and the second part is the high voltage (HV).}
    \label{fig:circuit}
\end{figure}

The electrical circuit is divided into two sections, a low voltage control region and a high voltage discharge circuit, as shown in Figure~\ref{fig:circuit}. The low voltage is transformed to high voltage using a Matsusada DC-DC converter ~\cite{ps}. On the low voltage side, a potentiometer is used to control the output voltage in a linear way, followed by a current amplification stage.
%I moved this into the next paragrah-RD
%On the high voltage side, a capacitor stores the energy and is periodically discharged through a thyristor to create a high voltage across the upper resistor that is higher than the breakdown voltage, leading to an arc. The thyristor was implemented as an improvement over insulated-gate bipolar transistor, since the latching behaviour of thyristors is well characterized and leads to a predictable behaviour. The thyristor is opened using two stacked Opto-triac devices to withstand the high voltage. 

The high voltage discharge system is a capacitor-discharge circuit that delivers controlled sparks. The energy is stored in a bank of capacitors across the electrode gap with a total capacitance of $\approx 100$~$\upmu$F. The energy stored can be released in a microsecond-scale pulse over a gap that is adjustable to change the intensity of the spark. The voltage on the electrodes is set with a resistive voltage divider. 

A thyristor (VS-30TPS12L) drives the lower-potential electrode momentarily to ground to discharge the energy stored in the capacitor through the gap between electrodes. This discharge leads to the arc in the chamber. This circuit configuration is limited to a maximum voltage of 1.2~kV. The thyristor was implemented as the latching behaviour of thyristors is well characterised with predictable behaviour, contrary to the unpredictable latching behaviour of an insulated-gate bipolar transistor. An external function generator sets the gate width and the repetition rate via two Opto-Triac devices assembled in series to withstand the high applied voltages, which open the thyristor. Additionally, the Opto-Triacs provide complete isolation of the high and low voltage sections.

\section{Light Spectrum of the Flash Lamp}
First, we validated the emission spectrum produced by the flash lamp. An Ocean Optics QEPro spectrometer (200~nm--1000~nm) recorded the light emission spectrum. For this test, the optical filter window was replaced with an acrylic window to allow both VUV and non-VUV light to reach the spectrometer. The resulting intensity curve was normalised, and then an algorithm was applied for peak identification using the NIST Atomic Spectra Database~\cite{nist}. The list is restricted to wavelengths between 200~nm and 1000~nm. We observed signals from the spectrometer associated with argon, carbon, hydrogen, nitrogen, and oxygen. Emissions from trace hydrocarbons, copper, aluminium, and iron originating from the chamber were considered negligible.

Each catalogue line was labelled with its element and ionisation stage (e.g., Ar I, Ar II) and the measured peaks were matched to the nearest reference line within a wavelength range of $\pm$ 0.2~nm. If multiple candidates were found inside this window, the match was classified as ambiguous. The four defined categories are:
\begin{itemize}
    \item Argon match: This peak corresponds to a wavelength peak of argon.
    \item Ambiguous with argon: There is an argon wavelength peak known within the tolerance of $\pm$ 0.2~nm at a wavelength shared with oxygen, nitrogen, and hydrogen. 
    \item Ambiguous without argon: There are multiple candidates within the tolerance of $\pm$ 0.2~nm, and none of them is associated with argon. 
    \item Non-argon match: This peak corresponds to another element.
\end{itemize}

\begin{figure}[H]
    \centering
    \includegraphics[width=0.85\linewidth]{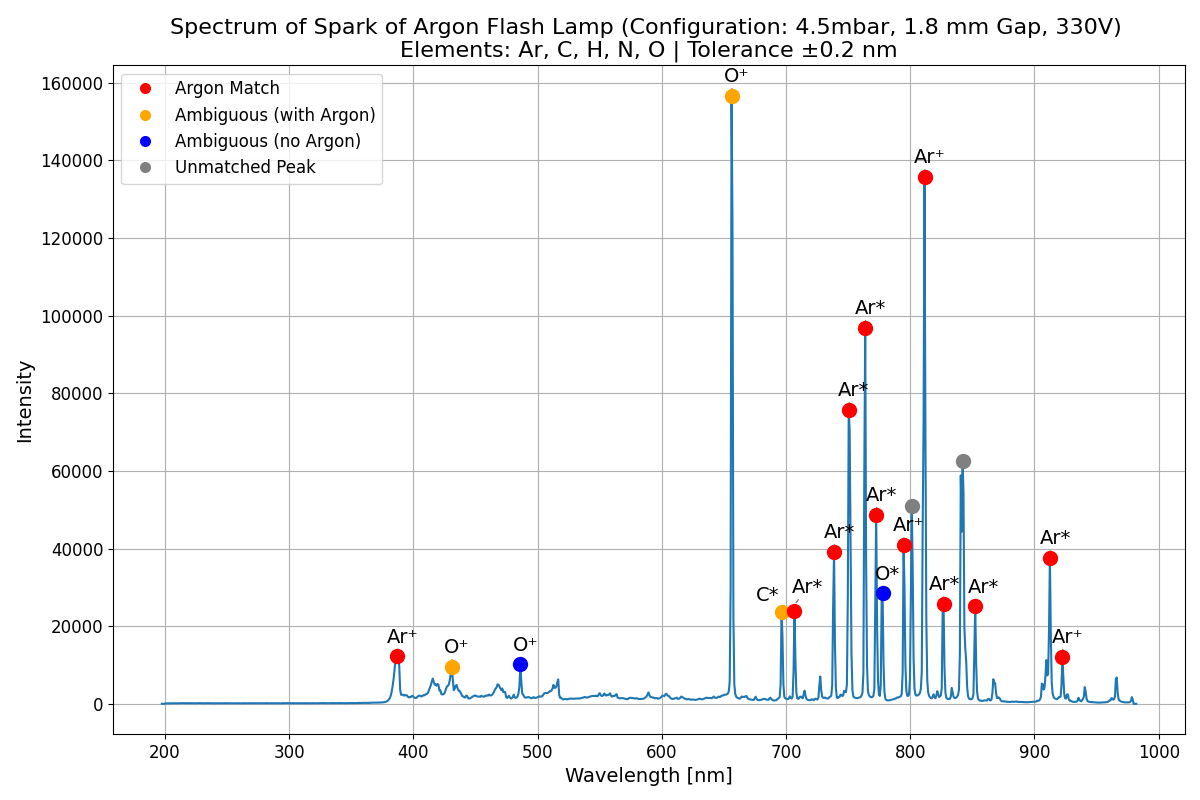}
    \caption{Spectrum of light collected over 2~s of pulsed sparking between two electrodes within the range of the spectrometer (200~nm--1000~nm). The electrodes were at a distance of 1.8~mm at a frequency of 10~Hz, while the chamber had argon at a pressure of approximately 450~Pa. The identified elements of the peaks are labelled.}
    \label{fig:wavelengthpeaks}
\end{figure}

Figure \ref{fig:wavelengthpeaks} shows the spectrum that was recorded over 2~s with a sparking frequency of 10~Hz, a spark gap of 1.8~mm, and a pressure of 450~Pa of argon. In total, for Figure \ref{fig:wavelengthpeaks}, 19 wavelength peaks were detected, with 12 assigned uniquely to argon, three determined as ambiguous with argon, two peaks classified as unassigned without an overlapping argon emission peak, and the remaining two peaks assigned to other elements. The dominance of argon lines confirms that the present configuration successfully reproduces discharges that produce light at wavelengths associated with argon.

\section{Determination of VUV Emission}
We performed a test to validate the device's applicability for testing SiPMs with VUV light. The spark was triggered across a distance of approximately 0.1~mm, and the filtering window was installed to transmit light at a central value of 125~nm $\pm$ 2.5~nm. The resulting distribution of wavelengths of transmitted light has a full width of 20~nm $\pm$ 5.0~nm. All the other wavelengths are filtered by a factor of $10^{-3}$--$10^{-4}$~\cite{optical_window}. The placement of the filter relative to the electrodes is shown in Figure~\ref{fig:chamber_front}. 

\begin{figure}[H]
    \centering
    \includegraphics[width=0.9\linewidth]{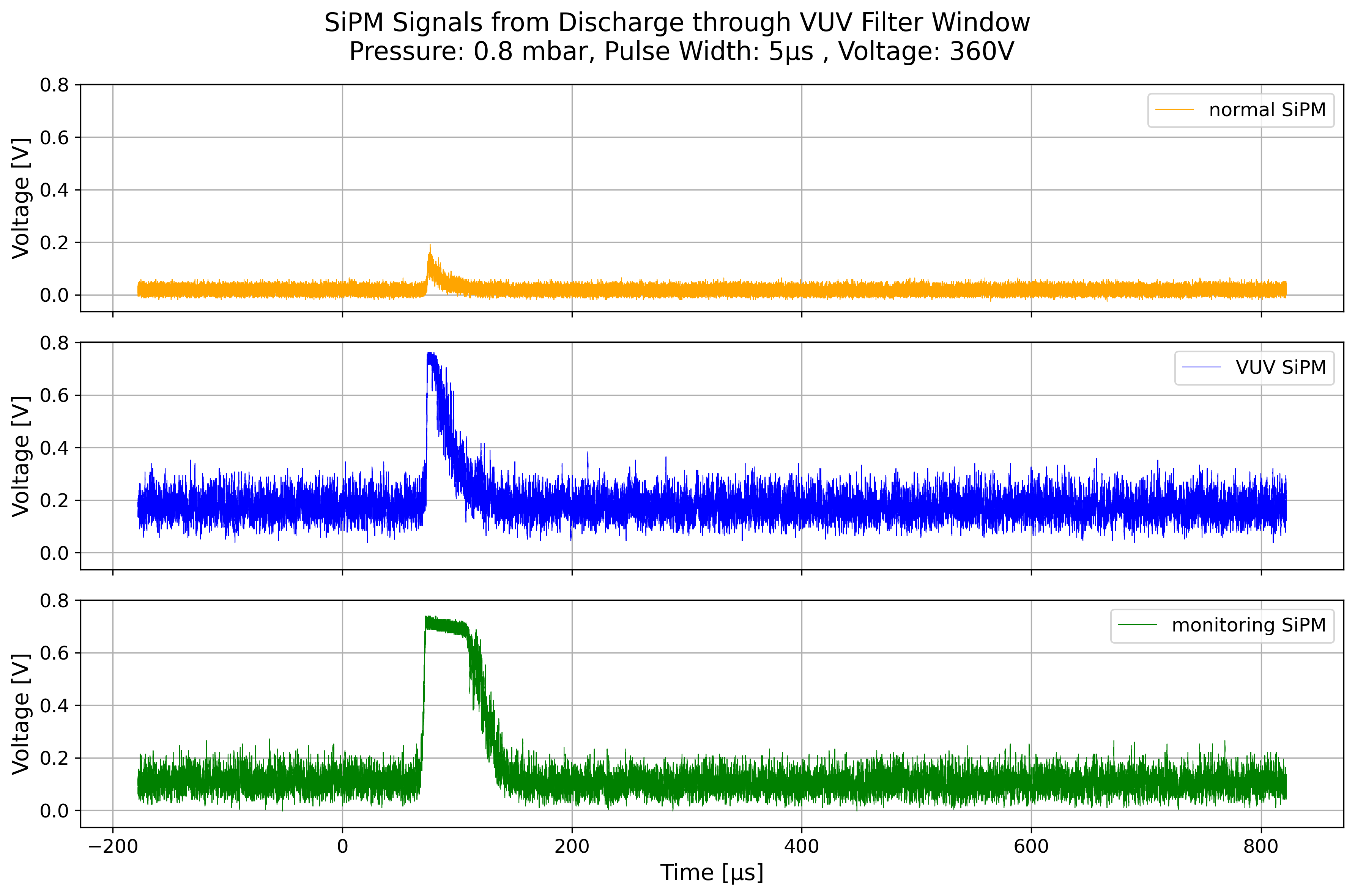}
    \caption{Measurements of light from the flash lamp with a filter in the filtering window for the normal and VUV-sensitive SiPMs. The SiPM optimised for visible light detected less light than the saturated VUV-sensitive SiPM.}
    \label{fig:m3}
\end{figure}

Two test SiPMs were mounted on the external side, a few millimetres from the window on the outside of the chamber. Of the two SiPMs mounted, one is a Hamamatsu SiPM (S13360-6050CS) that is sensitive to 270~nm to 900~nm, and the other is a Hamamatsu VUV-sensitive SiPM (S13370) that is sensitive to light with wavelengths from 120~nm to 900~nm. Both the standard and VUV-sensitive SiPMs were connected to an oscilloscope for the analysis of their waveforms. The monitoring SiPM built into the flash lamp chamber was also connected as a reference. The result of this measurement is shown in Figure~\ref{fig:m3}. As can be seen, signals on both SiPMs were observed in time with the spark creation. The VUV-sensitive SiPM observed a higher voltage output, indicating that it detected a higher signal peak when compared to the non-VUV-sensitive SiPM. We note that there is a volume of air between the chamber and the SiPM. The VUV light interacting with air may lead to spectral contamination from UV and visible light.

\begin{figure}[H]
    \centering
    \includegraphics[width=0.9\linewidth]{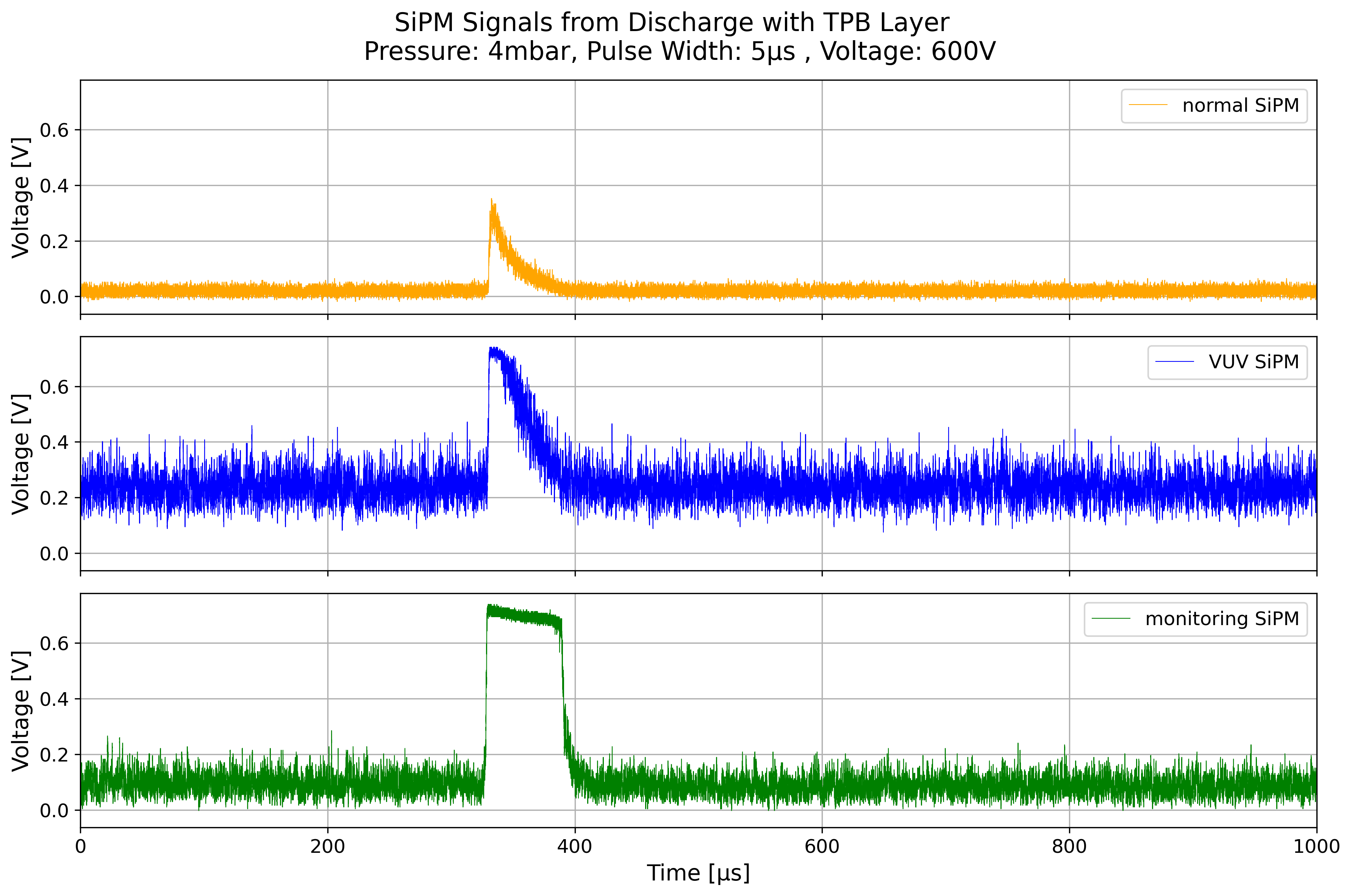}
    \caption{Measurements of light from the flash lamp. The filtering window was covered by a dichroic mirror with TPB to shift the produced light to the visible spectrum.}
    \label{fig:m5}
\end{figure}

\begin{figure}[H]
    \centering
    \includegraphics[width=0.9\linewidth]{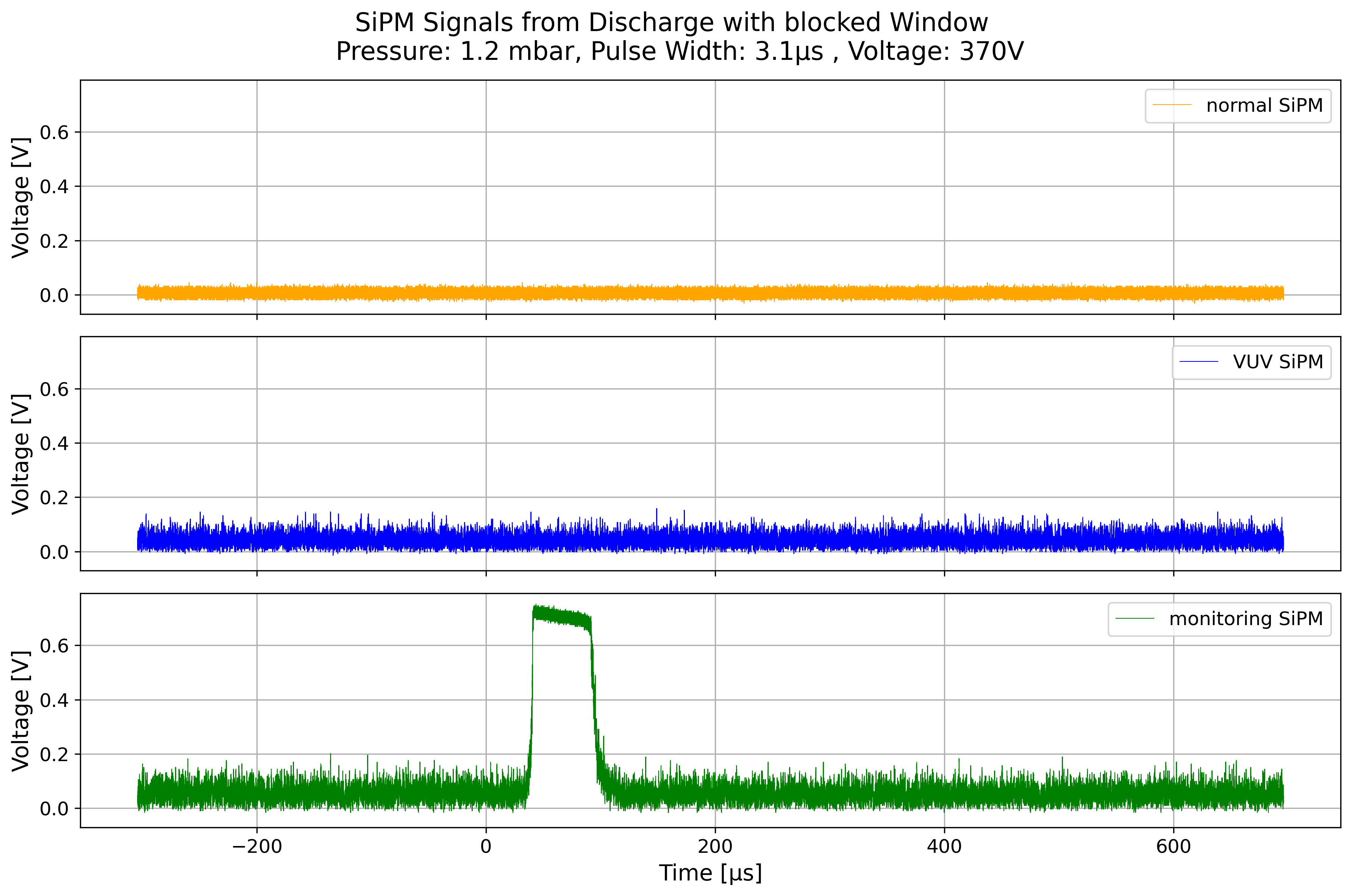}
    \caption{Measurements of light from the flash lamp. The filtering window was completely covered for this test, only allowing light to go to the monitoring SiPM.}
    \label{fig:m4}
\end{figure}

To confirm the filtering of extraneous light, an additional test was made with an added layer of wavelength-shifting TPB placed in front of the SiPMs. The TPB layer is exposed to the transmitted light from the filter and shifts the light to wavelengths corresponding to visible light. The ratio of light detected by the VUV SiPM and the normal SiPM before and after adding a layer of TPB was calculated. Integrating the waveform over the signal window, the normal SiPM detects roughly 3.1 times more light relative to the VUV SiPM when the TPB layer is added compared to without it. The resulting ratio between the SiPM signals indicates that the non-VUV-sensitive SiPM detected more light when including the wavelength shifter, as seen in Figure~\ref{fig:m5}, and provides evidence for the presence of VUV scintillation light in the transmitted light. A final test, shown in Figure~\ref{fig:m4},  blocking the entire window outside of the chamber was completed to confirm that signals observed by the SiPMs external to the flash lamp were due to the flash lamp and not due to noise.


\section{Conclusion}
We successfully built and tested a compact, flexible, and easy-to-operate VUV flash lamp. We demonstrated triggered discharges across an adjustable spark gap in argon gas. The measured emission spectrum shows good agreement with the known argon emission lines in the 200~nm--1000~nm range. Results indicate the production of VUV light by the flash lamp through cross-comparison measurements using a filtering window with a standard SiPM and a VUV-sensitive SiPM. %It was shown that it creates the intended argon emission spectrum on a triggered basis. 

To reduce the probability of light sensor saturation, future studies will investigate reducing the intensity of light produced by exploring different electrode gaps and applied voltages. Additionally, further investigations are required to understand how the air gap between the SiPMs and the chamber affects the spectrum of light reaching the sensors. Finally, the presented setup and measurements could be repeated with other noble elements, specifically xenon, to explore applications in other noble element-based particle physics experiments.

\acknowledgments{The research is funded by the Swiss National Science Foundation and by the Canton of Bern, Switzerland through grant 200020-204241. 
The study was supported by the Mechanic and Electronics Workshop at the Laboratory for High Energy Physics, Universität Bern, as well as the Albert Einstein Center for Fundamental Physics, Universität Bern.}

\conflictsofinterest{The authors declare no conflicts of interest.} 

%%%%%%%%%%%%%%%%%%%%%%%%%%%%%%%%%%%%%%%%%%
%\isPreprints{}{% This command is only used for ``preprints''.
\begin{adjustwidth}{-\extralength}{0cm}
%} % If the paper is ``preprints'', please uncommA critical requirement at the Laboratory for High-Energy Physics (LHEP) is a dependable, economical source of vacuum-ultraviolet (VUV) photons at 128 nm. In the Near-Detector Liquid-Argon Time-Projection Chamber (ND-LArTPC), neutrino interactions in liquid argon generate prompt scintillation via argon-excimer de-excitation, yielding photons almost exclusively at this wavelength (see Section 3.3.2). The light-readout system is therefore optimised for 128 nm detection.


\reftitle{References}

\bibliography{references}

\PublishersNote{}
%\isPreprints{}{% This command is only used for ``preprints''.
\end{adjustwidth}
\end{document}